# Lattice Boltzmann simulation of three-phase flows with moving contact lines on curved surfaces

Sheng Li[a], Yang Lu[a], Fei Jiang[b,c,d], Haihu Liu[a,*]

* E-mail: haihu.liu@mail.xjtu.edu.cn

[a] *School of Energy and Power Engineering, Xi'an Jiaotong University,*

*28 West Xianning Road, Xi'an 710049, China*

[b] *Department of Mechanical Engineering, Yamaguchi University, Ube 7558611, Japan*

[c] *Blue energy center for SGE technology (BEST), Yamaguchi University, Ube 7558611,*

*Japan*

[d] *International Institute for Carbon-Neutral Energy Research (WPI-I2CNER), Kyushu*

*University, Fukuoka 8190395, Japan*

A numerical method for simulating three-phase flows with moving contact lines on arbitrarily complex surfaces is developed in the framework of lattice Boltzmann method. In this method, the immiscible three-phase flow is modeled through a multiple-relaxation-time color-gradient model, which not only allows for a full range of interfacial tensions but also produces stable outcomes for a wide range of viscosity ratios. A characteristic line model is introduced to implement the wetting boundary condition, which is not only easy to implement but also able to handle arbitrarily complex boundaries with prescribed contact angles. The developed method is first validated by the simulation of a Janus droplet resting on a flat surface, a perfect Janus droplet deposited on a cylinder, and the capillary intrusion of ternary fluids for various viscosity ratios. It is then used to study a compound droplet subject to a uniform incoming flow passing through a multi-pillar structure, where three different values of surface wettability are considered. The simulated results show that the surface wettability has significant impact on the droplet dynamic behavior and final fluid distribution.

## I. INTRODUCTION

Three-phase flows with moving contact lines (MCLs) are encountered in nature and numerous industrial processes, such as Water-Alternating-Gas (WAG) injection for enhanced oil recovery [1,2], double emulsion production in microfluidics [3] and remediation of non-aqueous phase liquids (NAPLs) in groundwater [4,5]. In these processes, the fluids are confined in irregular geometries, causing very complex interface structures and movement as well as complex contact line dynamics, which have a vital influence on the behavior of three-phase flows. Take the enhanced oil recovery by the WAG injection as an example: the presence of interfacial tensions among oil, water and gas and MCLs on rock surfaces would determine if and when the displacement occurs, while influencing the displacement pattern and thus the oil recovery efficiency. Therefore, a deep understanding of three-phase flows with MCLs on irregular surfaces is of critical importance for the design and optimization of these processes and probably the related devices.

Numerical modeling and simulations have become a promising option to study three-phase flows with MCLs, which can provide detailed information about the flow field, such as pressure, velocity and fluid distributions. Traditionally, three-phase flows can be simulated by the volume-of-fluid method, level-set method and the phase-field method, and based on these methods, several numerical strategies have been developed to incorporate the MCLs [6-10]. However, most of them are only suited to dealing with flat walls, which severely restricts their applications in practice.

In addition, these traditional numerical methods suffer from some drawbacks in simulating multiphase flows. For example, the volume-of-fluid and level-set methods require either sophisticated interface reconstruction algorithms or unphysical reinitialization processes to represent the interface, and they have to introduce an empirical slip model in order that the stress singularities can be avoided at the MCLs [11]. The phase-field method uses an interface thickness far greater than its actual value, which often leads to unphysical dissolution of small droplets and mobility-dependent numerical results [12]. Microscopically, the interface between different fluids and the MCLs are due to interparticle interactions [13]. Thus, mesoscopic level models are expected to describe more accurately three-phase flows with MCLs, especially on complex curved surfaces.

The lattice Boltzmann method (LBM), as a mesoscopic method, has been developed into an alternative to traditional numerical methods for simulating complex fluid flow problems. It is a pseudo-molecular method based on particle distribution functions that evolve via mesoscopic kinetic equations and reproduces macroscopic behavior [14]. The LBM has many advantages over traditional numerical methods, such as the algorithm simplicity, intrinsic parallelism and the ease of dealing with complex geometric boundaries [14,15]. In addition, its mesoscopic nature can provide many advantages of molecular dynamics, making the LBM particularly useful for the simulation of multiphase flows with MCLs. A number of multiphase models have been developed in the LBM community, but only a few of them are concerned with three-phase flows with MCLs. Bala et al. [16] extended their ternary phase-field LBM to model the contact line dynamics, where the wetting boundary condition is implemented through two different methods, namely forcing and geometric methods. The basic idea of the forcing method is to add a forcing term related to the fluid composition at the fluid nodes adjacent to the solid wall, similar to the one used in the multicomponent pseudo-potential model, while the geometric method is to specify the values of the order parameter at the ghost nodes according to a geometrical wetting criterion. They found that the geometric method produces more accurate outcomes especially when the interpolation algorithm proposed by Lee and Kim [17] is used. Liang et al. [18] proposed a new wetting boundary scheme for wall-bounded ternary fluid flows in the framework of phase-field LBM. It was demonstrated that the proposed scheme not only can preserve the reduction consistency property of the diphasic case, but also is accurate for a wide range of contact angles. Zhang et al. [19] developed a multicomponent pseudo-potential model for three-phase flows in random porous media, in which the prescribed contact angles are achieved by adjusting the interaction parameters that control the strength of the interfacial tension between the solid and fluids. Nevertheless, the phase-field and pseudo-potential LBMs for three-phase flows have some inherent limitations: the former cannot conserve of mass for each fluid and yields mobility-dependent numerical results; the latter suffers from the lack of flexibility in adjusting interfacial tensions, high spurious velocities and poor stability in handling unequal fluid viscosities [20]. These limitations can be overcome by the use of color-gradient model, which has shown great success in the simulation of three-phase flows with MCLs, see Refs. [20,21]. In these color-gradient models, the geometric wetting boundary condition proposed by Zhang et al. [10] is commonly used. Nowadays, the implementation strategy of the geometric wetting boundary condition is presented only for a flat surface [20,21], and it remains unclear how to extend such a boundary condition to the case of curved surfaces.

In this work, we present a numerical method for modeling the three-phase flow with MCLs, particularly on arbitrarily curved surfaces. In this method, the immiscible three-phase flows are described by the LBM color-gradient model recently developed by Yu et al. [22], which is able to accurately simulate ternary fluid flows with a full range of interfacial tensions and ensure the conservation of mass for each fluid. To enhance the numerical stability of the model in dealing with large viscosity ratios and obtain viscosity-independent wall location, the multiple-relaxation-time (MRT) collision operator is adopted instead of its Bhatnagar-Gross-Krook counterpart [23-25]. A characteristic line model, along with a weighted contact angle model used for allocating the desired contact angle to each fluid [10], is introduced to implement the wetting boundary condition, which can handle arbitrarily complex boundaries with much ease. The capability and accuracy of this method are first tested through three benchmark cases, namely a Janus droplet resting on a flat surface, a perfect Janus droplet deposited on a cylinder, and the capillary intrusion of ternary fluids for a broad range of viscosity ratios. It is then used to simulate a compound droplet subject to a uniform incoming flow passing through a multi-pillar structure. We show how the surface wettability influences the droplet dynamic behavior and final fluid distribution.

## II. NUMERICAL MODEL

### A. Lattice Boltzmann color-gradient model for immiscible three-phase flows

The LBM color-gradient model recently proposed by Yu et al. [20,22] is used to simulate the immiscible three-phase flow for its capability and accuracy in simulating a full range of interfacial tensions. In this model, three different immiscible fluids, namely red, green and blue fluids, are considered, which are represented by the distribution functions $f_{\alpha,r}$, $f_{\alpha,g}$ and $f_{\alpha,b}$ respectively. The total distribution function is defined as $f_\alpha = \sum_k f_{\alpha,k}$ $(k = r,\ g\ \text{or}\ b)$, which undergoes a collision step as

$$f_\alpha^\dagger(\mathbf{x},t) = f_\alpha(\mathbf{x},t) + \Omega_\alpha(\mathbf{x},t) + \overline{F}_\alpha(\mathbf{x},t), \tag{1}$$

where $f_\alpha(\mathbf{x},t)$ is the total distribution function in the $\alpha$ th velocity direction at the position $\mathbf{x}$ and time $t$; $f_\alpha^\dagger$ is the post-collision distribution function; $\Omega_\alpha$ is the collision operator, and $\overline{F}_\alpha$ is the forcing term.

To obtain viscosity-independent wall location and enhance the stability of the model in dealing with high viscosity ratios, the MRT collision operator is adopted, which is given by [26,27]

$$\Omega_\alpha(\mathbf{x},t) = -\sum_\beta \left(\mathbf{M}^{-1}\mathbf{S}\mathbf{M}\right)_{\alpha\beta} \left[ f_\beta(\mathbf{x},t) - f_\beta^{eq}(\mathbf{x},t) \right], \tag{2}$$

where $\mathbf{M}$ is the transformation matrix and $\mathbf{S}$ is a diagonal relaxation matrix. $f_\alpha^{eq}$ is the equilibrium distribution function, and can be obtained by expanding the Maxwell-Boltzmann distribution in Taylor series of the local fluid velocity $\mathbf{u}$ up to the second order:

$$f_\alpha^{eq}(\mathbf{x},t) = \rho \omega_\alpha \left[ 1 + \frac{\mathbf{e}_\alpha \cdot \mathbf{u}}{c_s^2} + \frac{(\mathbf{e}_\alpha \cdot \mathbf{u})^2}{2c_s^4} - \frac{\mathbf{u} \cdot \mathbf{u}}{2c_s^2} \right], \tag{3}$$

where $\rho = \sum_k \rho_k$ is the total density with $\rho_k$ being the density of the fluid $k$; $c_s = c/\sqrt{3}$ is the speed of sound, with the lattice speed $c$ related to the lattice space $\delta_x$ and the time step $\delta_t$ by $c = \delta_x/\delta_t$ (for the sake of simplicity, $\delta_x = \delta_t = 1$ is used hereafter); $\omega_\alpha$ is the weight coefficient, and $\mathbf{e}_\alpha$ is the lattice velocity in the $\alpha$ th direction. For the D2Q9 lattice model used in this study, the lattice velocity $\mathbf{e}_\alpha$ is given by $\mathbf{e}_0 = (0,0)$, $\mathbf{e}_{1,3} = (\pm 1, 0)$, $\mathbf{e}_{2,4} = (0, \pm 1)$, $\mathbf{e}_{5,6} = (\pm 1, 1)$, $\mathbf{e}_{7,8} = (\mp 1, -1)$, and the weight coefficient $\omega_\alpha$ is given by $\omega_0 = 4/9$, $\omega_{1-4} = 1/9$, $\omega_{5-6} = 1/36$. The transformation matrix $\mathbf{M}$ is designed to linearly map the distribution functions onto the moment space, and is explicitly given by [26]

$$\mathbf{M} = \begin{pmatrix} 1 & 1 & 1 & 1 & 1 & 1 & 1 & 1 & 1 \\ -4 & -1 & -1 & -1 & -1 & 2 & 2 & 2 & 2 \\ 4 & -2 & -2 & -2 & -2 & 1 & 1 & 1 & 1 \\ 0 & 1 & 0 & -1 & 0 & 1 & -1 & -1 & 1 \\ 0 & -2 & 0 & 2 & 0 & 1 & -1 & -1 & 1 \\ 0 & 0 & 1 & 0 & -1 & 1 & 1 & -1 & -1 \\ 0 & 0 & -2 & 0 & 2 & 1 & 1 & -1 & -1 \\ 0 & 1 & -1 & 1 & -1 & 0 & 0 & 0 & 0 \\ 0 & 0 & 0 & 0 & 0 & 1 & -1 & 1 & -1 \end{pmatrix}. \qquad (4)$$

The diagonal relaxation matrix $\mathbf{S}$ reads as

$$\mathbf{S} = \text{diag}[0,\ s_1,\ s_2,\ 0,\ s_4,\ 0,\ s_6,\ s_7,\ s_8], \qquad (5)$$

with the relaxation rates $s_i$ given by

$$s_1 = s_2 = s_7 = s_8 = \frac{1}{\tau_f}, \quad s_4 = s_6 = \frac{8(2-s_1)}{8-s_1}, \qquad (6)$$

where $\tau_f$ is the dimensionless relaxation time related to the dynamic viscosity $\mu$ of the fluid mixture by $\mu = (\tau_f - 0.5)\rho c_s^2 \delta_t$. It should be noted that, for the sake of simplicity, the present method is limited to ternary fluids with equal densities, which could be extended to deal with unequal densities by using the techniques proposed in Ba et al. [28]. To allow for unequal viscosities for these fluids, a harmonic mean is adopted to determine the viscosity of the fluid mixture, i.e.,

$$\frac{\rho}{\mu} = \sum_k \frac{\rho_k}{\mu_k}, \qquad (7)$$

where $\mu_k$ is the dynamic viscosity of the pure fluid $k$.

The forcing term $\bar{F}_\alpha$ is introduced to generate the interfacial tensions between different fluids. In the MRT framework, the forcing term reads as [29]

$$\bar{F} = \mathbf{M}^{-1}\left(\mathbf{I} - \frac{1}{2}\mathbf{S}\right)\mathbf{M}\tilde{F}, \tag{8}$$

where $\mathbf{I}$ is a $9 \times 9$ unit matrix, $\bar{F} = [\bar{F}_0, \bar{F}_1, \bar{F}_2, ...\bar{F}_8]^T$, and $\tilde{F} = [\tilde{F}_0, \tilde{F}_1, \tilde{F}_2, ...\tilde{F}_8]^T$ with the element $\tilde{F}_\alpha$ given by [30]

$$\tilde{F}_\alpha = \omega_\alpha \left[\frac{\mathbf{e}_\alpha - \mathbf{u}}{c_s^2} + \frac{(\mathbf{e}_\alpha \cdot \mathbf{u})\mathbf{e}_\alpha}{c_s^4}\right] \cdot \mathbf{F}_s \delta_t. \tag{9}$$

In the above equation, $\mathbf{F}_s$ is the interfacial force, which contributes to the mixed interfacial region and is responsible for the local stress jump across the interface. According to the derivation of Yu et al. [22], the interfacial force can be expressed as

$$\mathbf{F}_s = \sum_k \sum_{l, l \neq k} \nabla \cdot \left[\frac{\sigma_{kl} A_{kl}}{2}|\mathbf{G}_{kl}|\left(\mathbf{I} - \mathbf{n}_{kl}\mathbf{n}_{kl}\right)\right], \tag{10}$$

where $\mathbf{G}_{kl} = C_l \nabla C_k - C_k \nabla C_l$ is the color gradient parameter and $C_k = \rho_k / \rho$ is the local mass fraction of the fluid $k$. $\sigma_{kl}$ is the interfacial tension between the fluid $k$ and the fluid $l$, and $\mathbf{n}_{kl} = \mathbf{G}_{kl} / |\mathbf{G}_{kl}|$ is the unit normal vector of the $k$-$l$ interface. $A_{kl}$ is a concentration factor controlling the activation of the interfacial tension at the $k$-$l$ interface, and it is given by $A_{kl} = \min\left(10^6 \rho_k \rho_l / \rho_k^0 \rho_l^0, 1\right)$, where $\rho_k^0$ is the density of the pure fluid $k$ [31]. The Chapman-Enskog expansion shows that the fluid velocity should be defined as

$$\rho \mathbf{u}(\mathbf{x}, t) = \sum_\alpha f_\alpha(\mathbf{x}, t)\mathbf{e}_\alpha + \frac{1}{2}\mathbf{F}_s(\mathbf{x}, t)\delta_t \tag{11}$$

to exactly recover the Navier-Stokes equations with a spatially varying body force.

In order to minimize the discretization errors, the fourth-order isotropic finite difference scheme is adopted to evaluate the partial derivatives in Eq. (10). Taking the variable $\psi$ as an example, its partial derivative can be calculated by

$$\frac{\partial \psi(\mathbf{x},t)}{\partial x_i} = \frac{1}{c_s^2} \sum_\alpha \omega_\alpha \psi(\mathbf{x}+\mathbf{e}_\alpha \delta_t, t) e_{\alpha i}. \tag{12}$$

Although the forcing term generates the interfacial tensions, it cannot guarantee the immiscibility of different fluids. In order to realize the phase segregation and maintain a reasonable interface, the recoloring algorithm proposed by Spencer et al. [32] is applied. Following their algorithm, for the fluid $k$, the recolored distribution function is given by

$$f_{\alpha,k}^{\dagger\dagger}(\mathbf{x},t) = \frac{\rho_k}{\rho} f_\alpha^\dagger(\mathbf{x},t) + \sum_{l, l\neq k} \beta_{kl} \omega_\alpha \frac{\rho_k \rho_l}{\rho} \mathbf{n}_{kl} \cdot \mathbf{e}_\alpha, \tag{13}$$

where $f_{\alpha,k}^{\dagger\dagger}$ is the recolored distribution function of the fluid $k$. $\beta_{kl}$ is a segregation parameter and is defined as [20]

$$\beta_{kl} = \beta^0 + \beta^0 \min\left(\frac{35\rho_r \rho_g \rho_b}{\rho^3}, 1\right) g(X_{kl}), \tag{14}$$

with

$$g(X_{kl}) = \begin{cases} 1, & X_{kl} < -1 \\ 1-\sin(\arccos(X_{kl})), & -1 \leq X_{kl} < 0 \\ \sin(\arccos(X_{kl}))-1, & 0 \leq X_{kl} \leq 1 \\ -1, & 1 < X_{kl} \end{cases}, \tag{15}$$

where $X_{kl} = (\sigma_{mk}^2 + \sigma_{ml}^2 - \sigma_{kl}^2)/(2\sigma_{mk}\sigma_{ml})$ and $\beta^0$ is the reference segregation parameter, which is fixed at 0.7 to be consistent with the segregation parameter in the two-phase color-gradient model [33,34]. Note that the use of segregation parameter defined by Eqs.(14) and (15) allows one to simulate three-phase flows with a full range of interfacial tensions and produce exactly the critical and near-critical states, as demonstrated by Yu et al. [20].

The recolored distribution functions then propagate to the neighboring lattice nodes, namely the propagation or streaming step, which reads as

$$f_{\alpha,k}(\mathbf{x}+\mathbf{e}_\alpha \delta_t, t+\delta_t) = f_{\alpha,k}^{\dagger\dagger}(\mathbf{x},t), \tag{16}$$

and the resulting distribution functions are used to compute the densities of three immiscible fluids, i.e., $\rho_k = \sum_\alpha f_{\alpha,k}$.

### B. Wetting boundary condition

For the three-phase flow with MCLs, the wettability of the solid surface can be described by three equilibrium contact angles, i.e, $\theta_{rg}$, $\theta_{br}$ and $\theta_{gb}$. According to the Young's equation, these contact angles are related to the interfacial tensions by

$$\begin{cases} \sigma_{rg} \cos\theta_{rg} = \sigma_{gs} - \sigma_{rs} \\ \sigma_{br} \cos\theta_{br} = \sigma_{rs} - \sigma_{bs} \\ \sigma_{gb} \cos\theta_{gb} = \sigma_{bs} - \sigma_{gs} \end{cases}, \tag{17}$$

where $\sigma_{ks}$ is the interfacial tension between the fluid $k$ and the solid surface. $\theta_{kl}$ represents the contact angle between the $k$-$l$ interface and the solid surface, which is measured from the side of the fluid $k$. Obviously, the contact angles $\theta_{kl}$ and $\theta_{lk}$ are a pair of complementary angles, i.e., $\theta_{lk} = \pi - \theta_{kl}$. In addition, one easily obtains the following relationship from Eq. (17), i.e.,

$$\sigma_{rg} \cos\theta_{rg} + \sigma_{br} \cos\theta_{br} + \sigma_{gb} \cos\theta_{gb} = 0, \tag{18}$$

which suggests that the interfacial tensions cannot be arbitrarily chosen when three contact angles are specified.

In the color-gradient LBMs, a popular way of enforcing wetting boundary condition is to specify the values of the mass fraction $C_k$ at the solid boundary nodes according to the desired contact angles, so that the related derivatives in Eq. (10) can be properly evaluated. Inspired by the work of Liu et al. [35], who dealt with two-phase flows with MCLs on complex solid surfaces, we introduce a characteristic line model to implement the wetting boundary condition in the present three-phase LBM.

To explain the characteristic line model, we first divide the lattice nodes into two categories, i.e.,

$C_F$: a list of the lattice nodes which are considered as fluid and

$C_S$: a list of the lattice nodes which are considered as solid.

Each of the two categories has two subcategories. For the lattice nodes in $C_F$, we define two subcategories, i.e.,

$C_{F_B}$: a list of the lattice nodes which belong to $C_F$ and are in contact with at least one lattice node in $C_S$ and

$C_{F_I}$: a list of the lattice nodes which belong to $C_F$ and are not in contact with any lattice node in $C_S$.

Similarly, we also define two subcategories for the lattice nodes in $C_S$, i.e.,

$C_{S_B}$: a list of the lattice nodes which belong to $C_S$ and are in contact with at least one lattice node in $C_F$ and

$C_{S_I}$: a list of the lattice nodes which belong to $C_S$ and are not in contact with any lattice node in $C_F$.

As an example, Figure 1 shows the categories and subcategories for the lattice nodes in the neighborhood of a circular solid grain, where only a part of solid grain is illustrated.

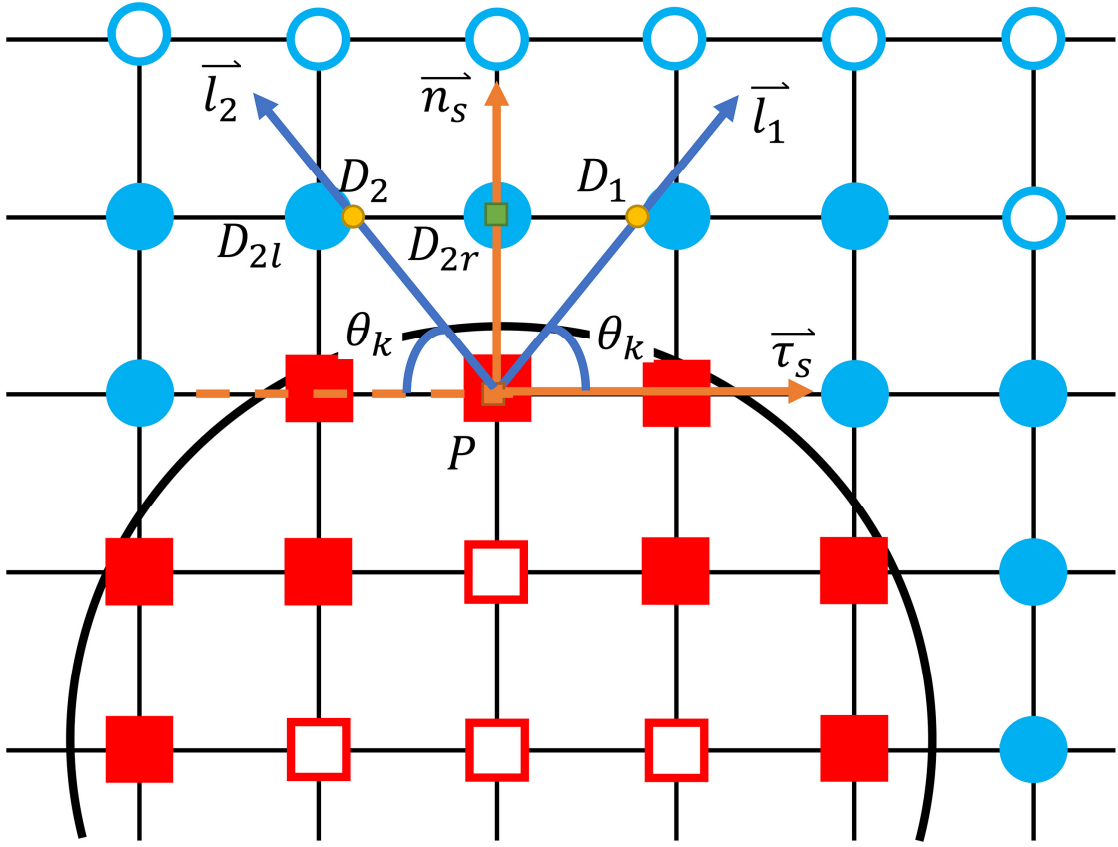

**FIG. 1.** Illustration of the categories and subcategories for the lattice nodes in neighborhood of a circular solid grain. Bold black lines represent the boundaries of solid grain. The small circles represent the lattice nodes in $C_F$, where the filled ones are in $C_{F_B}$ and the hollow ones are in $C_{F_I}$; the small squares represent the lattice nodes in $C_S$, where he filled one are in $C_{S_B}$ and the hollow ones are in $C_{S_I}$. $\mathbf{n_s}$ and $\mathbf{\tau_s}$ are the unit vectors normal and tangential to the solid surface at the node P; $l_1$ and $l_2$ are two characteristic lines originating from this node, which both make an included angle of $\left|\frac{\pi}{2}-\theta_k\right|$ with $\mathbf{n}_s$.

Then, we calculate the unit normal vector of the solid surface $\mathbf{n}_s$ for every lattice node in $C_{S_B}$. To suppress spurious velocities especially for a curved solid boundary, an eighth-order isotropic discretization scheme is adopted to evaluate $\mathbf{n}_s$, which reads as [36]

$$\mathbf{n}_s(\mathbf{x}) = \frac{\sum_\alpha \omega\left(|\mathbf{c}_\alpha|^2\right) s(\mathbf{x}+\mathbf{c}_\alpha \delta_t) \mathbf{c}_\alpha}{\left|\sum_\alpha \omega\left(|\mathbf{c}_\alpha|^2\right) s(\mathbf{x}+\mathbf{c}_\alpha \delta_t) \mathbf{c}_\alpha\right|}, \tag{19}$$

where $\mathbf{c}_\alpha$ is the $\alpha$ th mesoscopic velocity associated with the eighth-order isotropic discretization; $s(\mathbf{x})$ is an indicator function that equals 0 for $\mathbf{x} \in C_F$ and 1 for $\mathbf{x} \in C_S$; $\omega\left(|\mathbf{c}_\alpha|^2\right)$ is the eighth-order weight function and is given by

$$\omega\left(|\mathbf{c}_\alpha|^2\right) = \begin{cases} 4/21 & |\mathbf{c}_\alpha|^2 = 1 \\ 4/45 & |\mathbf{c}_\alpha|^2 = 2 \\ 1/60 & |\mathbf{c}_\alpha|^2 = 4 \\ 2/315 & |\mathbf{c}_\alpha|^2 = 5 \\ 1/5040 & |\mathbf{c}_\alpha|^2 = 8 \end{cases}. \tag{20}$$

In order to achieve a smooth transition of contact angles at the contact line, the weighted contact angle method is introduced to determine the local contact angles $\theta_k$, which are defined as [10]

$$\begin{cases} \theta_r = \dfrac{C_b}{1-C_r}\theta_{rb} + \dfrac{C_g}{1-C_r}\theta_{rg} \\ \theta_g = \dfrac{C_b}{1-C_g}\theta_{gb} + \dfrac{C_r}{1-C_g}\theta_{gr} \end{cases}. \tag{21}$$

Because $C_b + C_g + C_r = 1$, one only needs to enforce the wetting boundary condition for any two fluids, say red and green fluids here.

Next, we compute the values of $C_k$ in $C_{S_B}$ based on the obtained contact angles $\theta_k$. Taking the node P in Figure 1 as an example, one can obtain two characteristic lines originating from this node, i.e., $l_1$ and $l_2$, which make an included angle of $\left|\frac{\pi}{2}-\theta_k\right|$ with $\mathbf{n}_s$ and are symmetric with respect to $\mathbf{n}_s$. Extend the line $l_1$ until it intersects the grid line at $D_1$. If $D_1$ lies on a horizontal grid line, then we compute the value of $C_k$ at the point $D_1$, i.e. $C_{D_1}$, through a linear interpolation of $C_{lef}$ and $C_{rig}$: $C_{D_1} = C_{lef} + (x_{D_1} - x_{lef})(C_{rig} - C_{lef})$, where the subscripts 'lef' and 'rig' refer to the left and right grid nodes nearest to $D_1$. On the other hand, if $D_1$ lies on a vertical grid line, then we determine $C_{D_1}$ through $C_{D_1} = C_{low} + (x_{D_1} - x_{low})(C_{up} - C_{low})$, where the subscripts 'low' and 'up' refer to the lower and upper grid nodes nearest to $D_1$. In a similar way, we can obtain $C_{D_2}$ using the characteristic line $l_2$. It is noted that the intersection points $D_1$ and $D_2$ should satisfy the condition that their nearest grid nodes used for interpolation belong to $C_F$. If $D_1$ or $D_2$ cannot satisfy this condition, one should continue to extend the characteristic line so that an appropriate intersection point can be obtained.

After obtaining $C_{D_1}$ and $C_{D_2}$ we choose and assign one of them to $C_P$. In order to ensure the continuity of $C_k$ in $C_{S_B}$ along the tangential direction of the solid surface, we choose $C_P$ using the following discriminant function [35]

$$C_P = \begin{cases} \max\{C_{D1}, C_{D2}\} & \theta_k \leq 90°; \\ \min\{C_{D1}, C_{D2}\} & \theta_k > 90°. \end{cases} \quad (22)$$

Once the above treatment is applied to all the grid nodes in $C_{S_B}$, we are able to evaluate the derivatives in the interfacial force using Eq. (12). Thus, the desired contact angles are implicitly imposed by the gradient of $C_k$ in $C_{F_B}$.

## III. RESULTS AND DISCUSSION

In this section, three benchmark cases are first investigated to assess the accuracy and reliability of the present model for simulating three-phase flows with MCLs on flat and curved surfaces. The developed model is then used to simulate a compound droplet passing through a multi-pillar structure, and serval interesting phenomena are revealed.

### A. Janus droplet resting on a flat surface

Consider a semicircular Janus droplet, which consists of blue and red fluids and is surrounded by the green fluid, resting on a bottom wall in a $320 \times 180$ lattice domain. The initial radius of the Janus droplet is 80 lattices. The periodic boundary condition is used in the horizontal direction while no-slip boundary condition is imposed at the top and bottom walls through the halfway bounce-back scheme. In addition, the wetting boundary condition described in section II B is implemented at the top and bottom walls to achieve the desired contact angles.

Three fluids are assumed to have equal viscosities of $0.1$, and their interfacial intensions are set to $\sigma_{gb} = \sigma_{rg} = \sigma_{br} = \sigma = 0.01$. Six groups of contact angles that satisfy Eq. (18) are simulated, i.e. (a) $\theta_{rg} = 90°$, $\theta_{br} = 60°$, $\theta_{gb} = 120°$, (b) $\theta_{rg} = 60°$, $\theta_{br} = 90°$, $\theta_{gb} = 120°$, (c) $\theta_{rg} = 120°$, $\theta_{br} = 60°$, $\theta_{gb} = 90°$, (d)

$\theta_{rg} = 60°$, $\theta_{br} = 120°$, $\theta_{gb} = 90°$, (e) $\theta_{rg} = 90°$, $\theta_{br} = 90°$, $\theta_{gb} = 90°$, and (f) $\theta_{rg} = 90°$, $\theta_{br} = 120°$, $\theta_{gb} = 60°$. Each simulations is run until the Janus droplet reaches a steady shape, and the final shape of the Janus droplet is plotted in Figure 2. It is seen that the Janus droplet exhibits different shapes for six groups of contact angles. To quantify the accuracy of numerical results, we measure the equilibrium contact angles from the final droplet shapes and the obtained results are presented in Table I, where the analytical values of contact angles are also shown for comparison. It is clear that the simulated equilibrium contact angles are all in good agreement with the corresponding analytical solutions, with the maximum relative error as low as 3.0%. This suggests that the present wetting boundary condition is able to accurately deal with static contact angles on a flat surface.

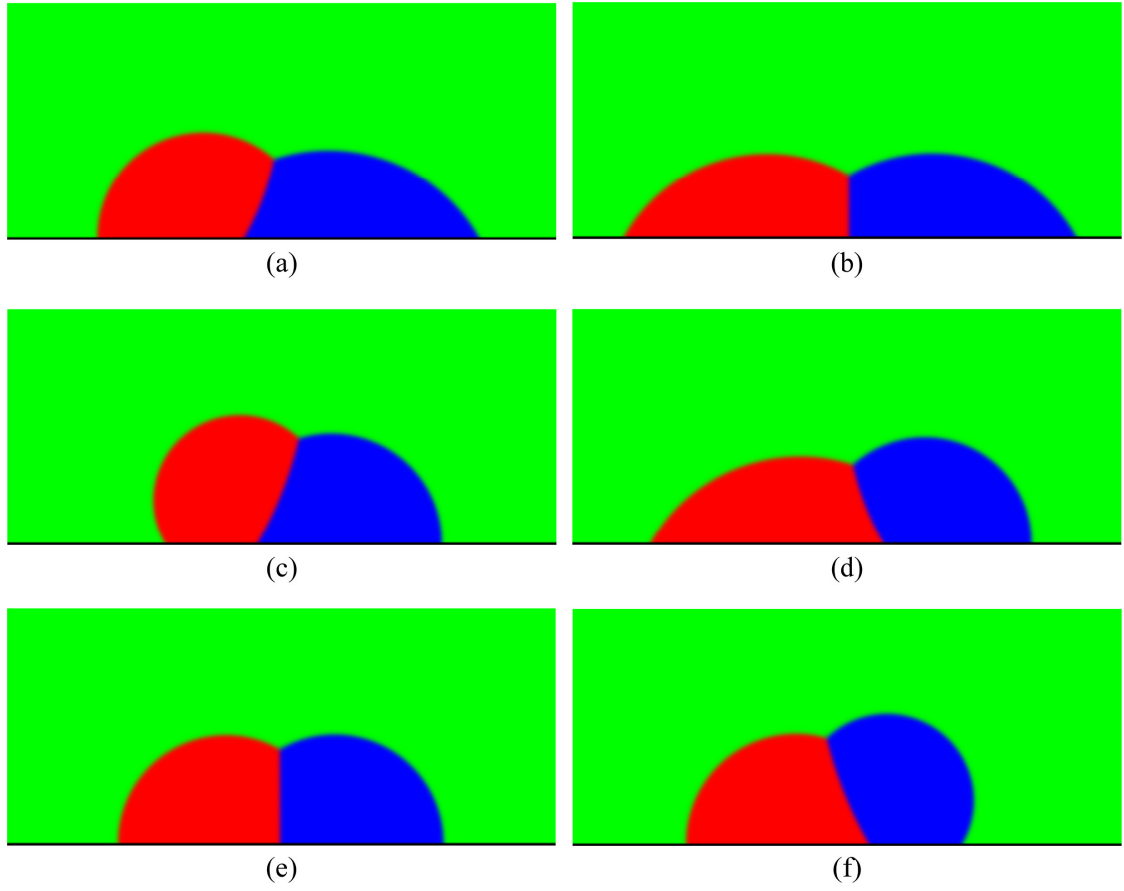

**FIG. 2.** The equilibrium shapes of a Janus droplet for different groups of static contact angles: (a) $\theta_{rg} = 90°$, $\theta_{br} = 60°$, $\theta_{gb} = 120°$, (b) $\theta_{rg} = 60°$, $\theta_{br} = 90°$, $\theta_{gb} = 120°$, (c) $\theta_{rg} = 120°$, $\theta_{br} = 60°$, $\theta_{gb} = 90°$, (d) $\theta_{rg} = 60°$, $\theta_{br} = 120°$, $\theta_{gb} = 90°$, (e) $\theta_{rg} = 90°$, $\theta_{br} = 90°$, $\theta_{gb} = 90°$, and (f) $\theta_{rg} = 90°$, $\theta_{br} = 120°$, $\theta_{gb} = 60°$.

**TABLE I.** Comparison between the simulated equilibrium contact angles and the corresponding analytical values for a Janus droplet resting on a flat surface.

|     |            | $\theta_{rg}$ | $\theta_{br}$ | $\theta_{gb}$ |
|-----|------------|--------|--------|--------|
| (a) | Analytical | 90°    | 60°    | 120°   |
|     | Simulated  | 89.1°  | 61.0°  | 120.0° |
| (b) | Analytical | 60°    | 90°    | 120°   |
|     | Simulated  | 59.7°  | 90.0°  | 119.9° |
| (c) | Analytical | 120°   | 60°    | 90°    |
|     | Simulated  | 118.2° | 60.8°  | 90.9°  |
| (d) | Analytical | 60°    | 120°   | 90°    |
|     | Simulated  | 60.0°  | 119.1° | 90.9°  |
| (e) | Analytical | 90°    | 90°    | 90°    |
|     | Simulated  | 89.1°  | 90.0°  | 90.9°  |
| (f) | Analytical | 90°    | 120°   | 60°    |
|     | Simulated  | 89.1°  | 119.2° | 61.8°  |

To determine the order of accuracy of the present method, we run the simulations with varying grid resolutions for the case of $\theta_{rg}=120°$, $\theta_{br}=60°$ and $\theta_{gb}=90°$. Figure 3 plots the relative error $E$ of the contact angle as a function of the initial radius $R$ of the Janus droplet, where the relative error is defined by the simulated contact angle $\theta_{rg}$ and the analytical value $\theta_{rg}^{A}$ as $E=\left|\theta_{rg}-\theta_{rg}^{A}\right|/\theta_{rg}^{A}\times100\%$. It can be seen that as the grid is refined, $\theta_{rg}$ approaches $\theta_{rg}^A$ approximately at a convergence rate of 1, suggesting that the present method is of first-order accuracy in the simulation of contact angles.

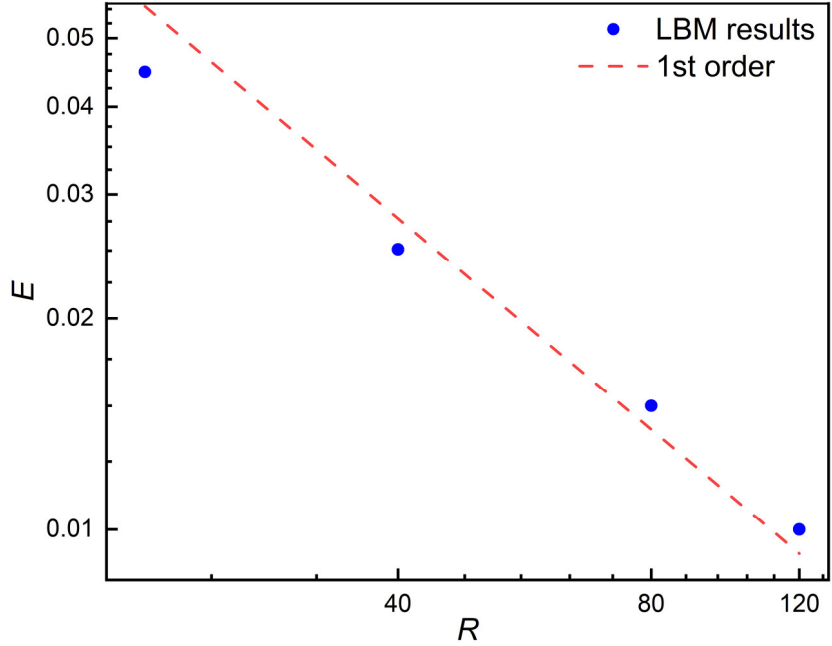

**FIG. 3.** The relative error ($E$) of the contact angle $\theta_{rg}$ as a function of the initial radius $R$ of the Janus droplet for the static contact angle test on a flat surface with $\theta_{rg}=120°$, $\theta_{br}=60°$, and $\theta_{gb}=90°$. Note that the discrete symbols represent the numerical results, while the dashed line corresponds to the first-order slope.

A numerical artifact observed in many multiphase solvers is the presence of spurious velocities at the phase interface, which is also true in the present method. Figure 4 plots the velocity fields for two typical cases, i.e., $\theta_{rg}=120°$, $\theta_{br}=60°$, $\theta_{gb}=90°$, and $\theta_{rg}=60°$, $\theta_{br}=120°$, $\theta_{gb}=90°$, in the final state. Obviously, spurious velocities are distributed in the entire computational domain and their maximum values ($|\mathbf{u}|_{max}$) are on the order of $O(10^{-4})$. On the basis of $|\mathbf{u}|_{max}$, we can calculate the spurious capillary numbers ($Ca = |\mathbf{u}|_{max} \mu / \sigma$) and find their values on the order of $O(10^{-3})$, which can be comparable to those in two-phase MCL problems [15]. In addition, we notice that higher spurious velocities are present not only in the vicinity of contact lines but also in the mixed region of three fluids, unlike in two-phase MCL case where spurious velocities are mainly present in the vicinity of contact lines.

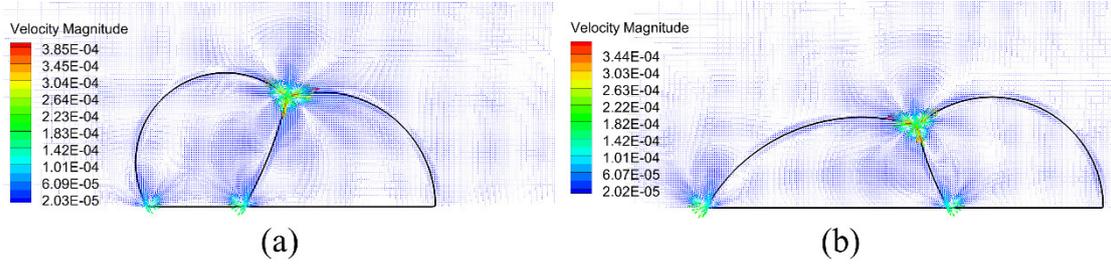

(a)  (b)

**FIG. 4.** Final velocity fields for a Janus droplet resting on a flat surface at (a) $\theta_{rg} = 120°$, $\theta_{br} = 60°$, $\theta_{gb} = 90°$ and (b) $\theta_{rg} = 60°$, $\theta_{br} = 120°$, $\theta_{gb} = 90°$.

Although the simulated contact angles match well with the analytical ones, it remains unknown whether the droplet morphology can be accurately captured by the present method. Thus, we make a quantitative comparison between the interface profiles obtained from present method and those from the existing LBMs [18,20]. Without loss of generality, two cases, i.e. $\theta_{rg} = 120°$, $\theta_{br} = 60°$, $\theta_{gb} = 90°$, and $\theta_{rg} = 60°$, $\theta_{br} = 120°$, $\theta_{gb} = 90°$ are considered, and the grid resolution is taken the same as used by Liang et al. [18], i.e. $R = 60$ and the domain size $L_x \times L_y = 300 \times 150$. Figure 5 shows the final shapes of the Janus droplet obtained by three different methods. It is clear that our simulation results are in excellent agreement with those obtained by Liang et al. [18] and Yu et al. [20], indicating that the present method is capable of describing the interfacial morphologies at the contact lines and in the mixed region of three fluids with satisfactory accuracy.

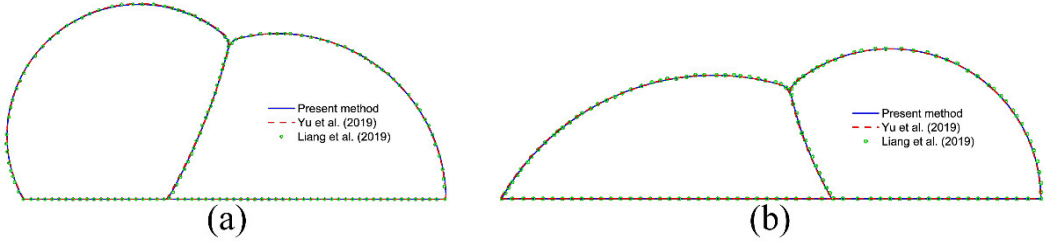

**FIG. 5.** Final shapes of a Janus droplet on a flat surface for (a) $\theta_{rg} = 120°$, $\theta_{br} = 60°$, $\theta_{gb} = 90°$, and (b) $\theta_{rg} = 60°$, $\theta_{br} = 120°$, $\theta_{gb} = 90°$. The simulation results obtained by Liang et al. [18] and Yu et al. [20] are also shown for comparison.

### B. Janus droplet deposited on a cylinder

To validate the capability of the present model in dealing with curved solid boundaries, we simulate a Janus droplet deposited on a cylinder. As depicted in Figure 6, a Janus droplet consisting of red and blue fluids is placed on a cylinder surface, which is immersed in the green fluid. To obtain the analytical solution of the problem, only the perfect Janus droplet is considered, which means that the interfacial tension between the red and blue fluids is negligibly small. Also, the red and blue fluids are assumed to have the same mass and physical properties, such as the density, viscosity and the surface wettability. Under this condition, the Janus droplet would eventually attain a circular arc shape, which can be theoretically determined using a simple geometric analysis (see below).

As shown in Figure 6, the droplet interface intersects with the solid surface at point $A$, and the points $O_j$ and $O_s$ are the centers of the Janus droplet and the cylinder, respectively. When the radius of Janus droplet ($R_j$), the radius of solid cylinder ($R_s$) and $\theta_{rg}$ are given, one can calculate the distance between the points $O_j$ and $O_s$, i.e., $D_{sj}$, using the law of cosine in $\triangle O_j A O_s$, which reads as

$$D_{sj} = R_s^2 + R_j^2 - 2R_s R_j \cos\theta_{rg}. \tag{23}$$

Thus, the equilibrium shape of the Janus droplet is obtained theoretically. In the simulations, $R_j$ and $R_s$ are fixed at 45 and 40 lattices, respectively. The cylinder is centered at $(100, 60)$ in a $200 \times 200$ lattice domain. The periodic boundary condition is used in the horizontal direction, while the halfway bounce-back scheme is imposed at the top and bottom walls to obtain the no-slip condition. Six groups of static contact angles are investigated: (a) $\theta_{rg} = \theta_{bg} = 30°$, (b) $\theta_{rg} = \theta_{bg} = 45°$, (c) $\theta_{rg} = \theta_{bg} = 60°$, (d) $\theta_{rg} = \theta_{bg} = 90°$, (e) $\theta_{rg} = \theta_{bg} = 120°$, and (f) $\theta_{rg} = \theta_{bg} = 150°$. The other parameters are chosen as $\sigma_{rg} = \sigma_{gb} = 0.01$ and $\mu_r = \mu_g = \mu_b = 0.1$. For each group of contact angles, the simulation is run until reaching the steady state, and the final results are recorded.

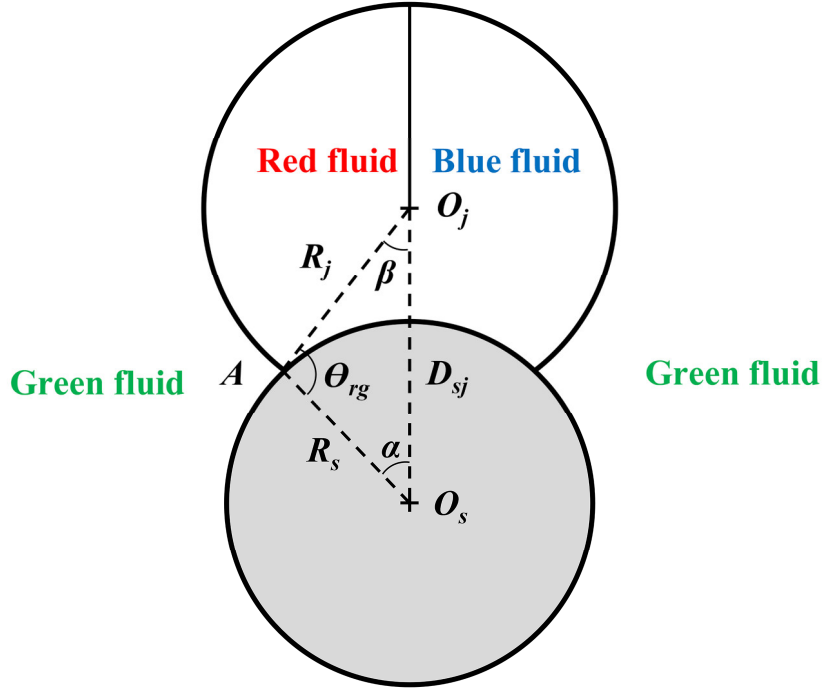

**FIG. 6.** Schematic diagram of a perfect Janus droplet resting on a cylinder.

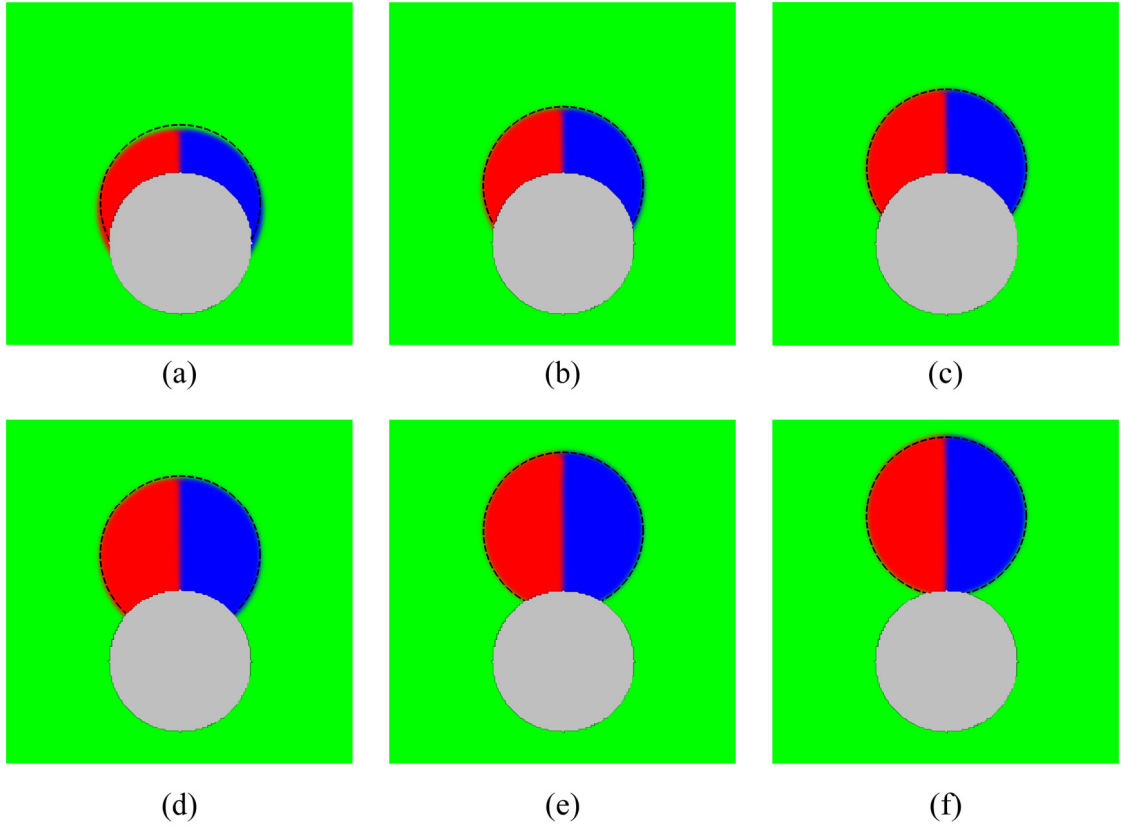

**FIG. 7.** A Janus droplet resting on a cylinder for various contact angles: (a) $\theta_{rg} = \theta_{bg} = 30°$, (b) $\theta_{rg} = \theta_{bg} = 45°$, (c) $\theta_{rg} = \theta_{bg} = 60°$, (d) $\theta_{rg} = \theta_{bg} = 90°$, (e) $\theta_{rg} = \theta_{bg} = 120°$, and (f) $\theta_{rg} = \theta_{bg} = 150°$. The black dotted lines represent the analytical droplet profiles, while the gray region represent the cylinder.

Figure 7 shows the comparison between the simulated Janus droplet shapes and analytical ones for various contact angles, in which the analytical solutions are represented by the black dotted lines. Obviously, the simulated results show good agreement with the analytical solutions except subtle difference observed in the case of $\theta_{rg} = \theta_{bg} = 30°$, which is attributed to relatively low grid resolution near the contact lines as highlighted by the circles in Figure 7(a). Such good agreement indicates that the present three-phase model can offer accurate prediction of static contact angles on curved solid surfaces, especially for the contact angles varying from $45°$ to $150°$. In addition, we notice in Figure 8 that, like in the flat surface cases, spurious velocities are still on the order of $O(10^{-4})$, and they are mainly distributed in the vicinity of contact lines and the mixed region of three fluids.

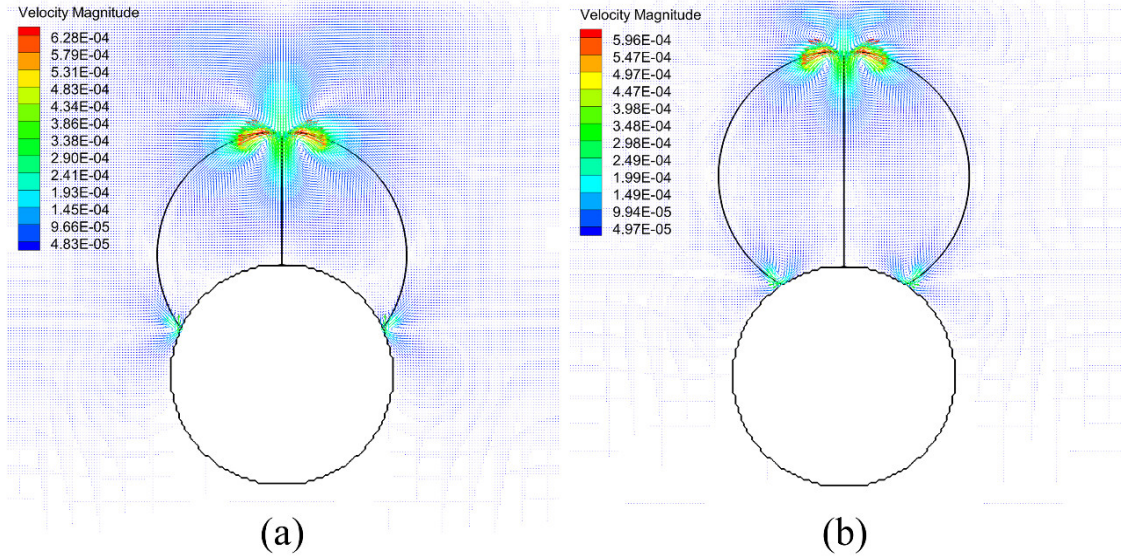

**FIG. 8.** Final velocity fields for a perfect Janus droplet resting on a cylinder at (a) $\theta_{rg} = \theta_{bg} = 60°$ and (b) $\theta_{rg} = \theta_{bg} = 120°$.

### C. Capillary intrusion of ternary fluids

Capillary intrusion has become a classic benchmark for examining whether a two-phase model is able to simulate MCLs problems since Washburn [37] derived its analytical solution. Motivated by the work of Washburn [37], Yu et al. [20] derived an analytical solution for the capillary intrusion of ternary fluids, which can be used to assess the present three-phase model. Figure 9 shows the setup for the capillary intrusion of ternary fluids, where a two-dimensional capillary tube with length $L$ and width $H$ is considered. Initially, the red, green and blue fluids are successively distributed in the capillary tube from the left to the right, and their initial lengths are $s_0$, $L_g$ and $L - s_0 - L_g$, respectively. The contact angles between the fluids and solid walls are $\theta_{gr}$ and $\theta_{gb}$. For the sake of simplicity, we focus on the spontaneous capillary intrusion in which the pressure difference between the inlet and outlet equals to 0.

According to the derivation of Yu et al., the intrusion length of the red fluid is governed by the following differential equation [20]

$$\frac{\mathrm{d}s(t)}{\mathrm{d}t} = \frac{H}{6} \frac{\sigma_{gb}\cos\theta_{gb} - \sigma_{gr}\cos\theta_{gr}}{\mu_r s(t) + \mu_g L_g + \mu_b \left[L - L_g - s(t)\right]}. \tag{24}$$

The above equation can be solved with the initial condition of $s(0) = s_0$, and the resulting analytical solution is

$$s(t) = \begin{cases} \dfrac{A}{B}t + s_0 & \mu_r = \mu_b; \\ s(t) = \dfrac{-B + \sqrt{B^2 - 2Cf(t)}}{C} & \mu_r \neq \mu_b \end{cases} \tag{25}$$

with

$$f(t) = -\frac{1}{2}Cs_0^2 - Bs_0 - At, \tag{26}$$

where $A = H\left(\sigma_{gb}\cos\theta_{gb} - \sigma_{gr}\cos\theta_{gr}\right)/6$, $B = \mu_g L_g + \mu_b\left(L - L_g\right)$, and $C = \mu_r - \mu_b$. It is worthy to note that in the analytical derivation of Eq.(25), the contact angles $\theta_{gb}$ and $\theta_{gr}$ are specified as wall boundary conditions. However, for comparing with simulations or experiments, they are better taken as the dynamic contact angles [38-40].

We run the simulations in a $450 \times 31$ lattice domain, which gives the size of capillary tube as $L \times H = 450 \times 30$. Three interfacial tensions are set to $\sigma_{gb} = \sigma_{rg} = \sigma_{br} = 0.01$, and two groups of contact angles are considered: (a) $\theta_{gr} = 60°$, $\theta_{gb} = 30°$ and (b) $\theta_{gr} = 60°$, $\theta_{gb} = 45°$. For each group of contact angles, three different viscosity ratios are simulated, i.e., $\mu_r : \mu_g : \mu_b = 1:1:1$, $\mu_r : \mu_g : \mu_b = 1:0.01:1$, and $\mu_r : \mu_g : \mu_b = 1:0.01:0.01$, which are achieved by varying $\mu_g$ and $\mu_b$ whilst keeping $\mu_r = 0.35$. In order to ensure zero pressure difference between the inlet and outlet, a modified periodic boundary condition is applied at the inlet and outlet, which can be implemented by exchanging the red and blue distribution functions pointing to the outside of the domain at the outlet (inlet) boundary before the propagation step. In addition, the halfway bounce-back is applied at the top and bottom walls, where the proposed wetting boundary condition is enforced as well to achieve the desired contact angles.

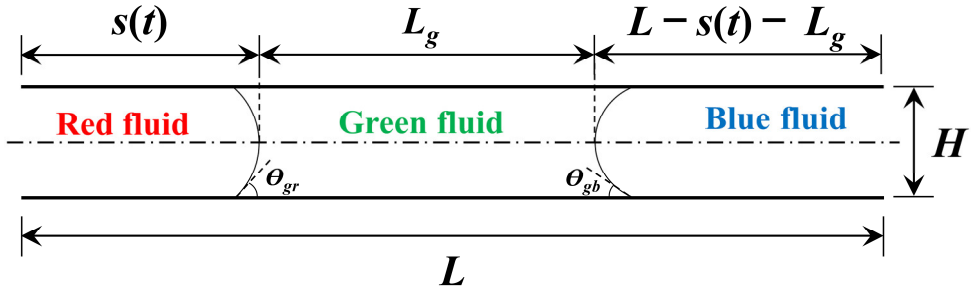

**FIG. 9.** The simulation setup for the capillary intrusion of ternary fluids. The capillary tube with length $L$ and width $H$, is connected to the reservoir of red fluid at the left inlet, and to the reservoir of blue fluid at the right outlet. At the time $t$, the $r-g$ and $g-b$ interfaces are located at $s(t)$ and $s(t) + L_g$, respectively.

Figure 10 compares the simulated results and the analytical predictions from Eq. (27) for different viscosity ratios at two groups of contact angles. Note that the analytical predictions are calculated using the dynamic contact angles measured from the simulated results, and the measured contact angles are very close to but slightly deviate from their prescribed values, with the maximum relative difference of 5.3%. It is clear that the simulated results agree well with the analytical predictions for a wide range of viscosity ratios and contact angles. This indicates that our model is able to provide satisfactory prediction of three-phase flows with MCLs for various viscosity ratios.

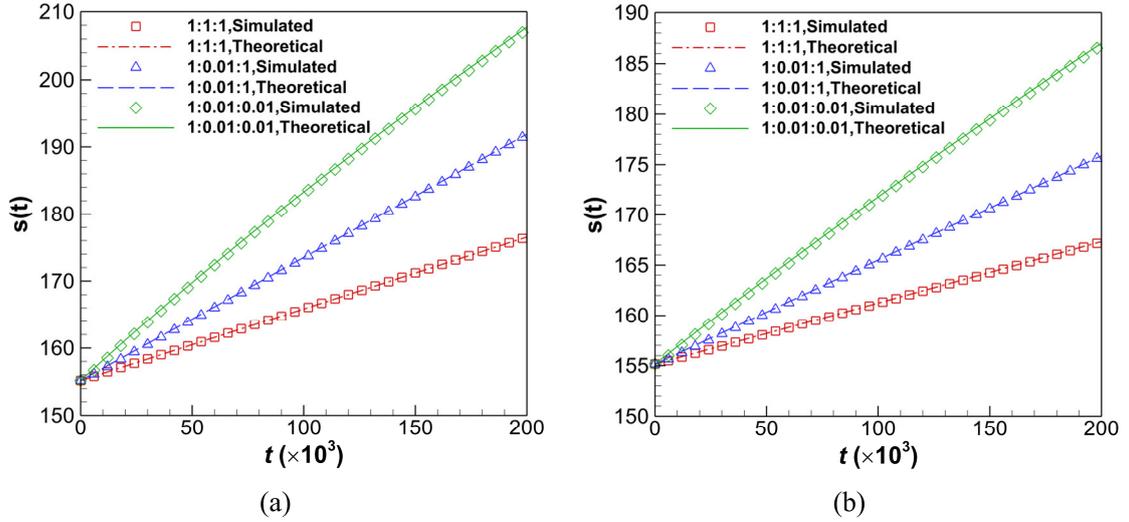

**FIG. 10.** The position of the $r-g$ interface versus time for three different viscosity ratios at (a) $\theta_{gr} = 60°$, $\theta_{gb} = 30°$ and (b) $\theta_{gr} = 60°$, $\theta_{gb} = 45°$. The hollow squares and circles represent the simulated results, while the solid lines are the corresponding theoretical solutions.

Finally, as an example, we choose the case of $\theta_{gr} = 60°$, $\theta_{gb} = 30°$ and $\mu_r : \mu_g : \mu_b = 1:1:1$ to show the velocity vectors throughout the capillary tube especially those close to the contact line regions, and the results are plotted in Figure 11. Note that the results of a two-phase capillary intrusion case, which considers only the green and blue fluids with $\theta_{gb} = 30°$, are also plotted for comparison. Clearly, the flow is one-directional and the velocity profile is parabolic everywhere except near the interface, confirming the Poiseuille flow assumption made in the derivation of Eq.(24). However, as seen from the local enlarged views, the flow appears to be distorted near the interface, which is necessary in order to allow the interface to advance with an uniform velocity along the vertical direction [41]. In addition, we notice that there is no evident difference in the velocity distribution between the two-phase and three-phase cases. This is understandable because the latter is essentially a combination of two two-phase interfaces, and there is no mixed region of three-phase fluids at all times.

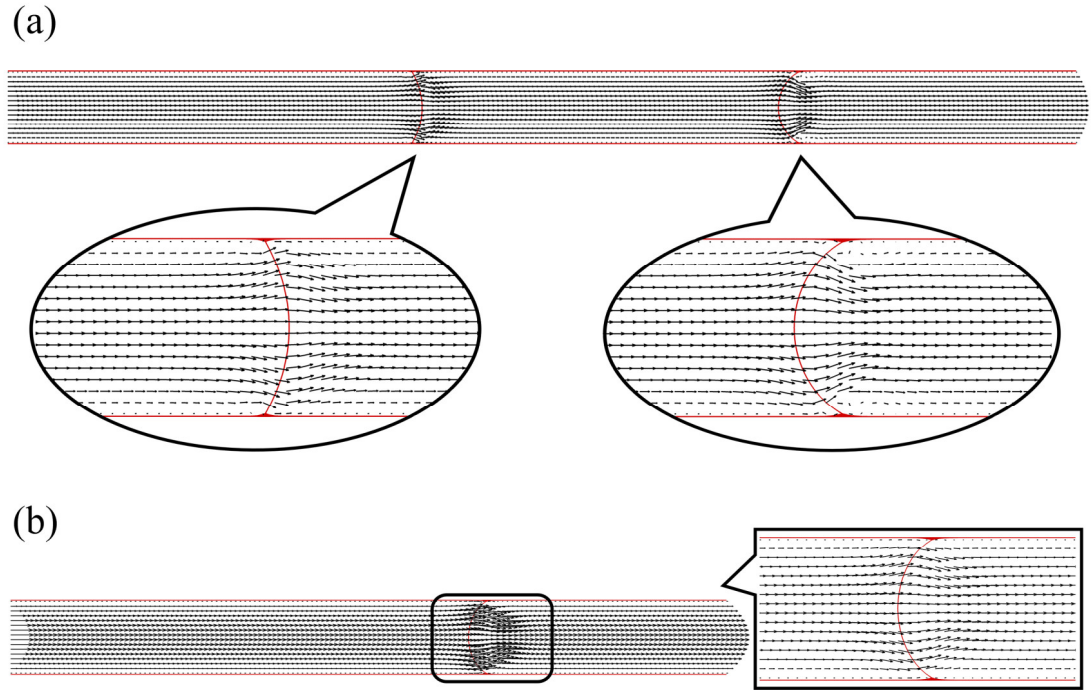

**FIG. 11.** Velocity vectors and the local enlarged drawings near the interface for (a) three-phase capillary intrusion with $\theta_{gr} = 60°$, $\theta_{gb} = 30°$ and $\mu_r : \mu_g : \mu_b = 1:1:1$, and (b) two-phase capillary intrusion involving green and blue fluids only. Note that the velocity vectors are shown every two grid nodes for better visualization.

From a sharp-interface point of view, the stress exhibits a non-integrable singularity

(known as the stress divergence) at the contact line when classical no-slip boundary condition is applied [42,43]. Like its two-phase counterpart [44,45], the present color-gradient LBM is of a diffuse-interface model, in which the contact line allows to move by virtue of diffusion so that the stress singularity can be overcome even when no-slip boundary condition is applied. In this sense, the present color-gradient LBM does not suffer from the problem of the stress singularity. This is also demonstrated by the velocity vectors near the moving contact lines, which are of finite magnitude and remain at a reasonable level.

### D. Compound droplet passing through a multi-pillar structure

To show the capability of the present model in dealing with dynamic problem with complex solid surfaces, we consider a compound droplet passing through a multi-pillar structure when subjected to a uniform incoming flow. The problem setup is depicted in Figure 12. Initially, a compound droplet, consisting of an outer (green) droplet with the radius $R_o = 50$ and an inner (red) droplet with the radius $R_i = 30$, is immersed in the third (blue) fluid and is placed upstream of five solid pillars, which are of the same radius, i.e. $R_s = 0.8 R_o$. The size of the entire computational domain is set to $L \times H = 16 R_o \times 7.5 R_o$. With the left bottom corner of the computational domain as the origin of the coordinate system, the compound droplet is initially centered at $(2R_o, 3.75R_o)$, and the five pillars are centered at $(6.9R_o, 3.75R_o)$, $(7.3R_o, 6.2R_o)$, $(7.5R_o, 1.6R_o)$, $(9.5R_o, 5.14R_o)$ and $(9.8R_o, 2.54R_o)$, respectively. A constant velocity $U$ is specified at the left inlet and a constant pressure is applied at the right outlet, with both implemented using the non-equilibrium bounce-back scheme proposed by Zou and He [46]. Periodic boundary conditions are used in the vertical direction. In this study, we choose the Reynolds number $Re = UR_o\rho_b / \mu_b = 0.75$, the capillary number $Ca = U\mu_b / \sigma_{bg} = 0.04$, and the viscosity ratio $\lambda = \mu_r : \mu_g : \mu_b = 1:1:1$, and investigate the effect of the surface wettability on the dynamic behavior of the compound droplet.

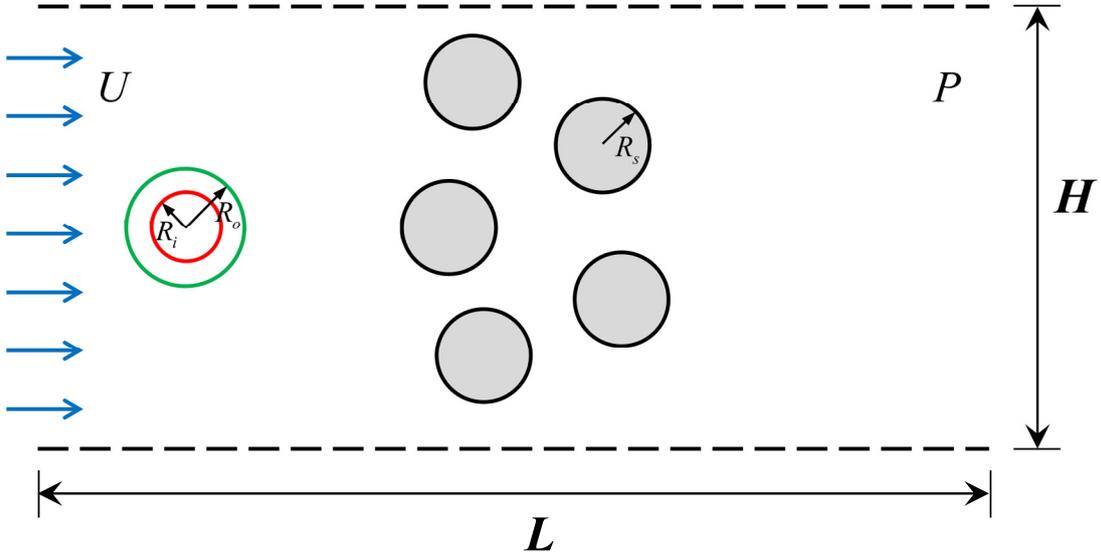

**FIG. 12.** A schematic diagram of a compound droplet passing through a multi-pillar structure when subjected to a uniform incoming flow.

We first consider the case of $\theta_{bg} = 135°$, $\theta_{br} = 45°$ and $\theta_{gr} = 40°$, which means that the solid surface is hydrophilic to the outer droplet but hydrophobic to the inner droplet. Figure 13 shows the time evolution of the shape and position of the compound droplet, where the dimensionless time is defined as $t^* = Ut / R_o$. At early time, the compound droplet is swept towards the pillars by the incoming flow and its shape does not change much. As the time evolves, the front of outer droplet gets in contact with the leftmost pillar and shows a tendency to wrap it (Figure 13b) because the solid surface is hydrophilic to the outer droplet. In addition, noticeable deformation is observed for both outer and inner droplets during this process. Under the influence of the viscous and capillary forces, the inner droplet then escapes from the tail of the outer droplet, as shown in Figure 13(c). Next, driven by the viscous shear flow from the blue fluid, the red droplet slides over the surface of the green droplet and passes through the passage between two neighboring pillars (Figure 13d). After that, the red droplet completely detaches from the green one and enters the upper flow pathway (Figure 13e). At the same time, the green droplet adheres to the entire solid pillar and keeps deforming. In particular, a liquid film forms around the upstream solid surface, which is gradually swept to the downstream of the pillar until disappears. At $t^* = 34.08$, the system has reached its equilibrium state, where the green droplet hangs from the leftmost pillar (Figure 13f). In addition, we interestingly observe that the red droplet also hangs from a solid pillar, even though the solid surface is hydrophobic to it. This phenomenon demonstrates that, even for a hydrophobic droplet, the capillary force from the solid pillar provides an adhesive force that anchors the droplet to the solid surface.

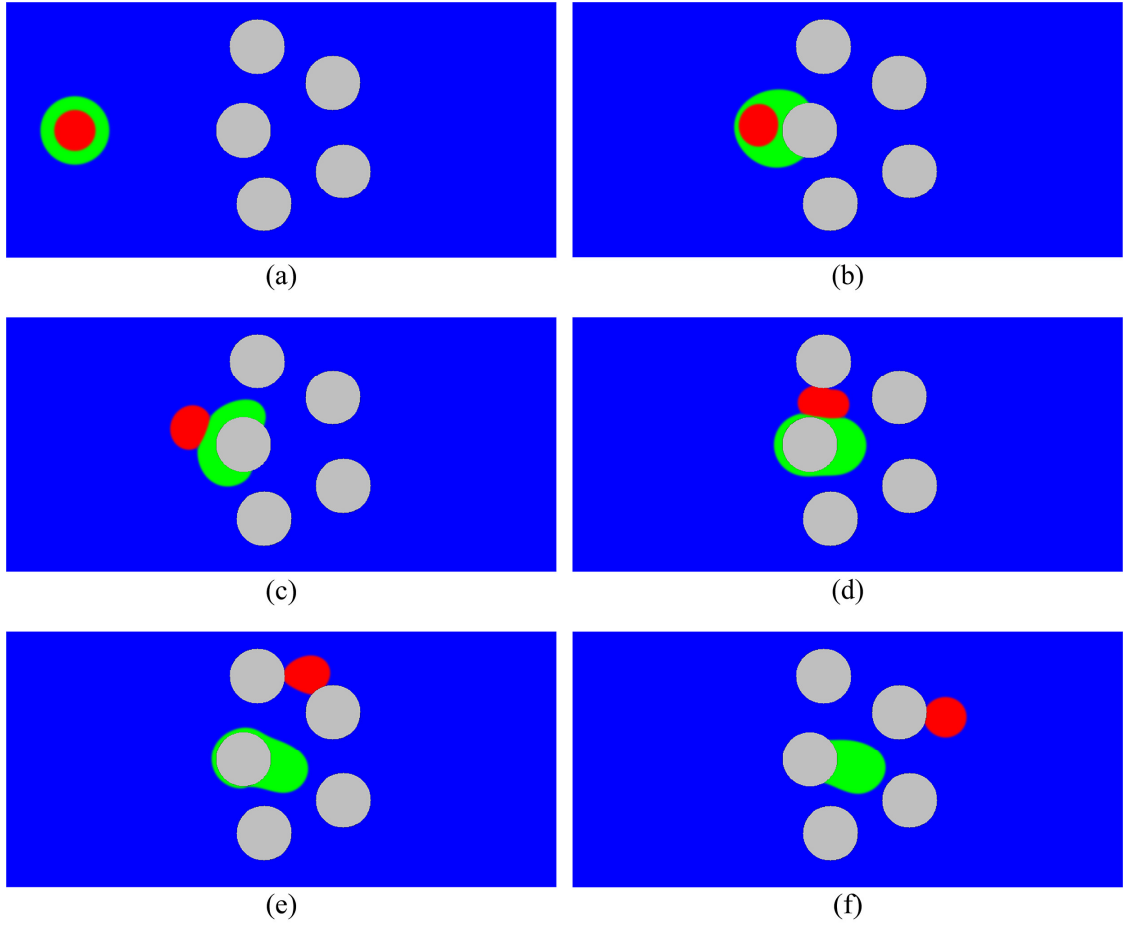

**FIG. 13.** Snapshots of the compound droplet for $\theta_{bg} = 135°$, $\theta_{br} = 45°$ and $\theta_{gr} = 40°$ at the times: (a) $t^* = 0$, (b) $t^* = 6.5$, (c) $t^* = 7.43$, (d) $t^* = 9.91$, (e) $t^* = 13.63$, and (f) $t^* = 34.08$. The gray regions represent the solid pillars.

We then turn to the case of $\theta_{bg} = 45°$, $\theta_{br} = 135°$ and $\theta_{gr} = 140°$, and the results are shown in Figure 14. Compared to the last case, the compound droplet exhibits different behaviors. First, once the front of the compound droplet touches the leftmost pillar, the inner droplet immediately breaks through the outer one (Figure 14b). Then, the outer droplet enters the flow passage between the leftmost pillar and the upmost pillar, and the inner droplet slides over its surface until adheres to the upmost pillar (Figure 14c). Contrary to the last case, the inner (red) droplet as the wetting fluid tends to wrap the solid pillar and eventually hangs from the upmost pillar (Figure 14d-f). On the other hand, the outer (green) droplet as the non-wetting fluid, passes through the right two solid pillars and finally moves downstream (Figure 14e-f). Such a fate of the outer droplet could be explained as follows. The outer droplet has a size almost twice that of the inner droplet, and it is swept into the middle flow passage in which the flow rate is higher. This would lead to a larger viscous force exerted on the outer droplet. When the viscous force is greater than the capillary force from the solid pillar, the outer droplet would pass through the multi-pillar structure and move downstream.

Finally, we simulate the case of $\theta_{bg} = 90°$, $\theta_{br} = 90°$ and $\theta_{gr} = 90°$, which means that the solid surface exhibits neutral wettability to all fluids. Figure 15 shows the corresponding snapshots of the moving compound droplet. At early time, like in the above cases, the inner droplet escapes from the tail of the outer droplet and both of them are swept into the upper flow passage, as can be seen from Figure 15(a)-(c). Then, the inner and outer droplets are temporarily trapped in the pore formed by the surrounding pillars (Figure 15d). Under the influence of viscous force and capillary force, they ultimately separate from each other. After that, the inner and outer droplets are swept into the upper and middle flow passages, respectively (Figure 15e). When the system reaches the equilibrium state, the inner (red) and outer (green) droplets adhere to two adjacent solid pillars and block the upper and middle flow passages, respectively (Figure 15f). Such a steady state is attainable because the capillary force from the solid pillar is strong enough to resist against the viscous force from the blue fluid. In addition, we notice that there is no liquid film formed around the solid pillar in this case, which is due to that the solid surface is of equal affinity to all fluids. Clearly, the droplet behavior and final fluid distribution are different from those in the above two cases, suggesting that the surface wettability is one of important factors governing this problem.

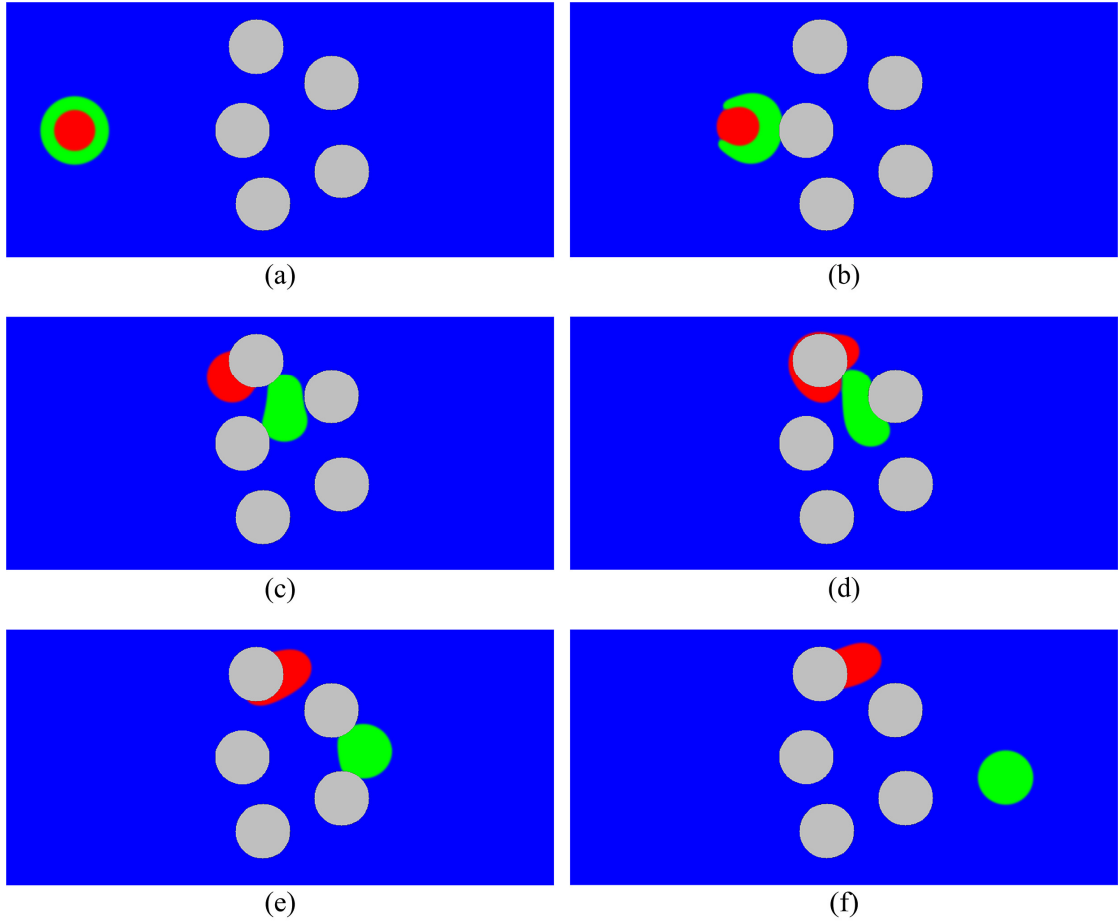

**FIG. 14.** Snapshots of the compound droplet for $\theta_{bg} = 45°$, $\theta_{br} = 135°$ and $\theta_{gr} = 140°$ at the times: (a) $t^* = 0$, (b) $t^* = 6.5$, (c) $t^* = 12.39$, (d) $t^* = 17.35$, (e) $t^* = 22.46$, and (f) $t^* = 27.89$. The gray regions represent the solid pillars.

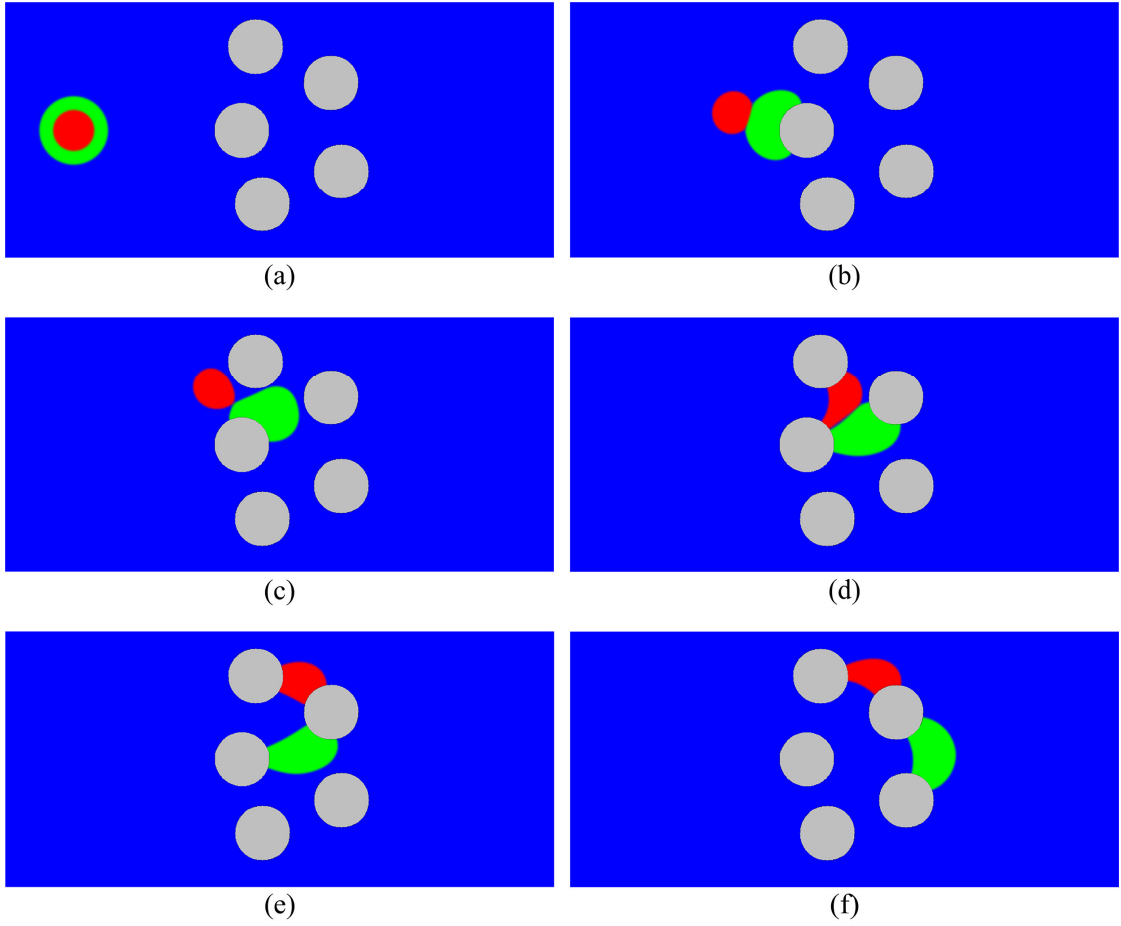

**FIG. 15.** Snapshots of the moving compound droplet for $\theta_{bg} = 90°$, $\theta_{br} = 90°$ and $\theta_{gr} = 90°$ at the times: (a) $t^* = 0$, (b) $t^* = 7.43$, (c) $t^* = 10.53$, (d) $t^* = 15.49$, (e) $t^* = 24.79$, and (f) $t^* = 37.81$. The gray regions represent the solid pillars.

## IV. CONCLUSIONS

A LBM is presented to simulate three-phase flows with MCLs on arbitrarily complex surfaces. In this method, the immiscible ternary fluid flow is modeled through a multiple-relaxation-time color-gradient model [22], which not only allows for a full range of interfacial tensions but also can handle a wide range of viscosity ratios. A characteristic line model is introduced to implement the wetting boundary condition on arbitrarily complex surfaces, in which the prescribed contact angles are realized by specifying the values of mass fraction at the solid nodes adjacent to the boundary using two possible characteristic lines. The capability and accuracy of the present method are tested through three benchmark cases, namely a Janus droplet resting on a flat surface, a perfect Janus droplet deposited on a cylinder, and the capillary intrusion of ternary fluids for a broad range of viscosity ratios. All the simulated results show good agreement with the corresponding analytical solutions, demonstrating that the present method can accurately simulate static and dynamic problems with various contact angles for a wide range of viscosity ratios. It is then used to simulate a compound droplet subject to a uniform incoming flow passing through a multi-pillar structure, in which complex fluid-solid interactions are involved. Three numerical cases with different surface wettabilities are simulated, and the results show that the droplet dynamic behavior and final fluid distribution can be significantly influenced by the surface wettability.

**ACKNOWLEDGMENTS**

This work is supported by the National Natural Science Foundation of China (Grant Nos. 12072257, 51876170), the National Key Project (Grant No. GJXM92579), the Natural Science Basic Research Plan in Shaanxi Province of China (Grant No. 2019JM-343), and JSPS through a Grant-in-Aid for Young Scientists (Grant No. 19K15100).